\documentstyle{ioplppt}                
\begin{document}
\title{Measures of the galaxy clustering}[Measures of the galaxy
clustering]

\author{Vicent J. Martinez}

\address{Departament d'Astronomia i Astrofisica, Universitat de
Val\`encia, 46100--Burjassot, Val\`encia, Spain}

\begin{abstract}
A brief introduction is given to some aspects of the statistical
description of the luminous matter distribution. I review the
features of the redshift surveys that arise in
the statistical analysis of the galaxy clustering.
Special topics include intensity functions, correlation
functions, correlation integrals, multifractals and multiscaling.

\end{abstract}

%
%

\section{Introduction}
I am sure that, in spite of the title of the School, at the end of these
two weeks, we shall have a less dark picture of the three-dimensional
distribution of matter in the Universe. There are still a huge amount
of unsolved problems regarding the origin and evolution of the
observed large scale structure in the Universe. Although important
developments have occurred during the last two decades, the task has
revealed so elusive, that most of the students of this School will
have interesting research projects on these topics in the following
years.

The statistical study of the clustering patterns formed by the
three-dimensional distribution of galaxies is one of the most
important observational clues to learn about the physical
processes that led to the large scale structure of the Universe.
A detailed statistical description of the observed distribution
of matter in the Universe is needed to confront theoretical
predictions of models of structure
formation, such as $N$-body simulations involving dark matter,
against  observations.

The aim of this lecture is to introduce some statistical aspects of
the description of the clustering in the Universe. This
introduction will be followed by more detailed lectures given by Drs.
Borgani and Coles.

\section{Surveys of galaxy redshifts}

In the 1980s different groups of astronomers started systematic
observational programs to construct a true three-dimensional
fair sample of the Universe. The task consists in measuring the location
in space of galaxies lying in the studied region of the sky. In
addition to its angular position, we need to know the distance to each
object. This is usually done by measuring the redshift $z$ in the
spectrum of the galaxies, which is related to the line-of-sight
recession velocity, $v_{\mbox{\tiny \rm rec}} = cz$.
The Hubble law permits to
estimate the distance $d$ to each galaxy, as $v_{\mbox{\tiny \rm rec}}
\simeq v_H = Hd$ (but see sec. 2.3), and therefore
to have the three-dimensional map of the Universe.

We denote by $\{ \vec x_i \}_{i=1}^N$ the position of the $N$ galaxies
in a portion of the universe with volume $V$. Techniques of point
fields statistics may be used to describe the statistical features of
the distribution of such galaxies.
Some words of caution might be given before
embarking on the different techniques of the statistical analysis.
Catalogues of galaxies are not simple point samples as could be many
of the planar point processes usually studied in the literature of
spatial statistics (like the positions of trees in forests). To handle
properly galaxy samples, we need to have in mind some of the
characteristics of their construction \cite{Feig92}.

\subsection{Galaxy obscuration}
The extragalactic optical light does not reach the Earth uniformly
from all directions. The plane of the Milky Way, our own Galaxy, is
filled with interstellar dust which absorbs most of the light coming
from extragalactic sources. Therefore catalogues are incomplete below
galactic latitudes $|b| < 20^{\circ} - 30^{\circ}$. This fact implies
that the band of the sky corresponding to low galactic latitudes
is usually not considered in the optical samples analyzed. The
geometry of the three-dimensional regions often becomes irregular
because of this and other observational constraints. E.g.,
the analysis of the CfA-I \cite{Huc83} redshift survey is usually
performed in the Northern Galactic hemisphere (with $b \ge
40^{\circ}$) and equatorial declination $\delta > 0^{\circ}$. Other
kind of obscurations are related to the observational devices is the
mask of the {\it IRAS} satellite \cite{qdot,Mar94}.
The QDOT-{\it IRAS} redshift survey
covers 74 \% of the sky after removal of the masked regions and
the band corresponding to galactic latitude $|b| \leq 10^{\circ}$
({\it IRAS} looks in the infrared and safely identifies galaxies
closer to the Galactic plane than optically observed galaxies).

The brightness of the galaxies is also affected by the galactic
absorption. This effect is usually modelled by the cosecant law in
latitude $\Delta m = A \; \mbox{\rm cosec} \; b$.

\subsection{Brightness and apparent magnitude limit}

Galaxies in a redshift survey have different intrinsic brightness.
Most of the catalogues are built by fixing an apparent magnitude limit
$m_{\lim}$. Therefore galaxies with $m > m_{\lim}$ are not seen by the
telescope or are not considered because of observational strategies. An
apparent magnitude-limited sample is therefore not uniform in
space, as intrinsically faint objects are only seen if they are
close enough to the Earth. To analyze this kind of flux-limited
samples we can follow two strategies:
\begin{enumerate}
\item
One can extract volume-limited samples by fixing a value of the depth
$D_{\max}$ (in $h^{-1}$ Mpc, $h$ being the Hubble constant in units
of 100 Mpc$^{-1}$ km s$^{-1}$) and keeping only galaxies brighter than
\begin{equation}
M_{\lim} = m_{\lim} - 25 - 5 \log (D_{\max}).
\end{equation}
For example, if the catalogue has an apparent magnitude limit
$m_{\lim} = 14.5$ and we consider as the maximum depth of the volume to
be studied $D_{\max} = 100 \, h^{-1}$ Mpc, only galaxies with absolute
magnitude $M \le -20.5 + 5 \log h$ will remain in the volume-limited
sample. With this strategy, however, we loose a part of the information
provided by the redshift survey. To avoid this problem we can
follow the second strategy.

\item
The second procedure is based on the knowledge of the
{\sl selection function}: $\varphi(x)$. The function $\varphi(x)$
gives an estimate of the probability that a
galaxy more brilliant than a given luminosity cutoff,
at a distance $x$ is included in the sample. If  the sample is
complete up to a distance $R$, $\varphi(x) = 1$ for
$x \le R$. For example, in Fig. 1, we show the selection
function of one QDOT-{\it IRAS} subsample for
galaxies with $L \ge 10^{9.19} \, L_{\odot}$.
With this luminosity limit, the sample is complete up to
$R = 40 \, h^{-1}$ Mpc \cite{Mar93}. Beyond this distance the
selection function $\varphi (x)$ falls down and attains values
close to zero for $R \simeq 200 \, h^{-1}$ Mpc.

The selection function is derived from the {\sl luminosity function}
$\phi (L)$. The luminosity function is defined by requiring that
the mean number of galaxies, per unit
volume, with luminosity in the range $L$ to $L+ dL$, is
$\phi(L) dL$. The empirical luminosity function is often
fitted by the analytical expression \cite{Sch76}
\begin{equation}
\phi(L) dL = \phi_{\ast} \left ({L \over L_{\ast}} \right )^{\alpha}
\exp \left (- {L \over L_{\ast}} \right ) d \left ( {L \over
L_{\ast}} \right ) ,
\end{equation}
where $L_{\ast}$ and $\alpha$ are the fitting parameters, while
$\phi_{\ast}$ is related with the number density of galaxies.
The previous expression in terms of magnitudes is
\begin{equation}
\phi(M) dM = A \,  \phi_{\ast} \left (
10^{0.4(M_{\ast}-M)} \right )^{\alpha+1}
\exp \left ( -10^{0.4(M_{\ast}-M)} \right ) dM \, .
\end{equation}
where $A= {2 \over 5} \ln (10)$.
Therefore, the selection function is just the ratio
\begin{equation}
\varphi(x) = {\int_{-\infty}^{M(x)} \phi (M) dM \over
\int_{-\infty}^{M_{\max}} \phi (M) dM} =
{\Gamma (\alpha+1, 10^{0.4(M_{\ast}-M(x))}) \over
\Gamma (\alpha+1, 10^{0.4(M_{\ast}-M_{\max})})} \, ,
\end{equation}
where $M(x) = m_{\lim} -25 - 5\log(x)$, $\Gamma$ being now the
incomplete
Gamma function, $M_{\max}=\max(M(x),M_{\mbox{\tiny \rm com}})$
and $M_{\mbox{\tiny \rm com}}$ is the absolute magnitude for which the
catalogue is complete.  The parameters of the luminosity function
depend on the sample \cite{lumpar}. Typical values are $\alpha =
-1.1$ and $M_{\ast}=-19.3 + 5\log h$.

Within this strategy, we can assign to each galaxy a weight $w =
1/\varphi(x)$ depending on its distance $x$ to us.
\end{enumerate}

\subsection{Redshift-space distortions}
The observed recession velocity $v_{\mbox{\tiny \rm rec}}$ is not only
due to the Hubble expansion.
Other components have to be added to $v_H$ to obtain
$v_{\mbox{\tiny \rm rec}}$.
Bulk flows or streaming motions on large scale or local
velocities within clusters on small scales might not be negligible.
The peculiar velocity is the velocity of a galaxy with respect to
the Hubble flow. Let us indicate its component along the line-of-sight
by $v_{\mbox{\tiny \rm pec}}$; then the observed recession
velocity is
\begin{equation}
v_{\mbox{\tiny \rm rec}} = cz = H_0 d + v_{\mbox{\tiny \rm pec}},
\end{equation}
and therefore we have to distinguish between `redshift space' and
`real space'; the first one is artificially produced by setting
each galaxy at the distance $d$ obtained by considering
$v_{\mbox{\tiny \rm pec}}=0$, and is therefore
a distorted representation of the second one. The effect
of this radial distortion is clearly illustrated when
dense clusters of galaxies, almost spherical in real space, appear
as structures elongated along the line of sight, in redshift space.
These structures are known as `fingers of God'.

\subsection{Segregation}
The point field formed by the galaxies in the surveyed volume is
clearly a {\sl marked point field}, in the sense that qualitative
marks like morphological type and quantitative marks like intrinsic
luminosity, distinguish different objects within the same catalogue.
The statistical properties of the spatial distributions of different
kinds of objects can be different. In fact, it is well established
that elliptical galaxies are more frequent in denser regions such as
rich clusters, while spirals are more often found in low-density
environments \cite{morse}. Statistical descriptors such as the
two-point correlation function or multifractal measures might provide
us with different results if they are applied on different categories
of galaxies. Although less evident, it seems also established
that a certain degree of luminosity segregation exists
\cite{luseg}, at least
for galaxies with absolute magnitude $M_B \le -20 + 5\log h$.
Bright galaxies are stronger correlated that faint galaxies. It seems
that both kinds of segregation exist but are independent effects.
Mass segregation will be reviewed in Dr Campos lecture during this
School. The segregation mechanisms are to be understood on the basis
of a convincing structure formation theory.

\subsection{Ergodic hypothesis}
Obviously all the statistical measures will be applied to a portion of
the Universe. Let $D$ be the size of a portion and $L$ be the
scale to which the measure refers. If $L$ is not much smaller
than $D$, and we apply the same measure to another portion of
size $D$, we expect to find different results.
This is sometimes referred to as `cosmic variance'. If $D \gg L$,
we shall have many realizations of the probability distribution
within our sample, and therefore we expect
that the results do not depend too much upon the studied region.
Statisticians would say that we are assuming {\sl ergodicity}
\cite{Peb80,Sto94}, in the sense that our sample is enough to
obtain statistically reliable results \cite{Buc93}, as it contains
an adequate number of independent realizations.

\section{Statistical measures}

We have summarized in the previous section  how the observation of the
Universe at large scale provide us with obscured, truncated, distorted
and segregated samples of galaxies. In spite of all their shortcomings,
they are of extraordinary interest in Cosmology. From the detailed
analysis of these samples we learn about the past and future of
the Universe.

In this section, we will introduce some of the mathematical techniques
often used to statistically describe the distribution of galaxies.

\subsection{Second-order characteristics}
I will start this section by using the terminology and notation
often employed by spatial statisticians and I will relate it with
that used by cosmologists.

In a point process, the galaxy distribution in our case, we can define
\cite{Sto94,Dig83} the { \sl second order intensity function}
$\lambda_2 (\vec{x_1},\vec{x_2})$ as follows: Let us consider two
infinitesimally small spheres centered in $\vec{x_1}$ and $\vec{x_2}$
with volumes $dV_1$ and $dV_2$. The joint probability that in each of
the spheres lies a point of the point process is approximately
\begin{equation}
dP = \lambda_2 (\vec{x_1},\vec{x_2})  dV_1 dV_2.
\end{equation}
(See \cite{Mar93,Dig83} for an exact definition). If the point
field is {\sl homogeneous} (sometimes called {\sl stationary}),
$\lambda_2 (\vec{x_1},\vec{x_2})$ depends only on the distance
$r=|\vec{x_1}-\vec{x_2}|$ and the direction of the line passing
through $\vec{x_1}$ and $\vec{x_2}$, $0 \le \beta < \pi$:
$\lambda_2 (r, \beta)$. If, in addition, the process is isotropic,
the angle $\beta$ becomes unimportant and the function depends only
on $r$, $\lambda_2 (r)$. In the following, we will assume the
Cosmological Principle: the Universe at large scales is homogeneous
and isotropic. Let $n$ be the mean number density of galaxies in a
huge volume, assumed to be a fair sample of the Universe.
The {\sl two-point correlation function} commonly used in Cosmology is
\cite{Peb80}
\begin{equation}
\xi(r) = {\lambda_2 (r) \over n^2} -1  .
\end{equation}

The expected number of points within a distance $r$ from an arbitrary
given galaxy is
\begin{equation}
\langle N \rangle_r = \int_0^r 4 \pi n s^2 (1 +\xi (s)) ds =
{4 \pi \over n} \int_0^r s^2 \lambda_2 (s) ds. \label{coin}
\end{equation}

The last expression may also be referred to as a correlation
integral $C(r)$ \cite{Mar95}. $K(r)=C(r)/n$ is called
the Ripley's $K$-function \cite{Rip81} and is
extensively used in the literature of point fields. In this context
$n$ is the first-order intensity function $\lambda$ of the point field,
which is constant for homogeneous processes.

\subsection{Estimators of $\xi(r)$}
Different estimators may be used to evaluate $\xi(r)$. For
volume-limited samples all the galaxies have the same weights $w=1$,
while when selection functions are used to account for the
incompleteness, each galaxy is counted with a weight $w \ge 1$.
Davis \& Peebles \cite{Dav83} use the estimator
\begin{equation}
1 + \xi_{DP}(r) = {DD(r) \over DR(r)}{N_R \over N_D} \, ,
\end{equation}
where $DD(r)$ is the number of pairs with separation $r$ in the galaxy
catalogue with $N_D$ galaxies and $DR(r)$ is the number of pairs with
separation $r$ between the data and a random distributed sample with
$N_R$ points.
Equivalently we can use \cite{Mar93,Riv86} the following estimator by
averaging over the $N$ galaxies of the sample
\begin{equation}
1 + \xi_R (r) = {1 \over N} \sum_{i=1}^N {N_i (r) \over n V_i(r)} \, ,
\end{equation}
where $N_i(r)$ is the number of galaxies lying in a shell of thickness
$dr$ at distance $r$ from galaxy $i$ and $V_i(r)$ is the volume of the
shell lying within the sample boundaries.

There is enough confidence in the power-law behaviour of the observed
galaxy two-point correlation function in the range of scales
$0.1 < r < 10 \, h^{-1}$ Mpc \cite{Mar93,Dav83,Dav88}.
\begin{equation}
\xi_{gg} (r) = \left ( {r \over r_g} \right )^{-\gamma} \, ,
\label{power}
\end{equation}
where the exponent $\gamma \simeq 1.8$ and the correlation length $r_g
\simeq 5 \, h^{-1}$ Mpc. It is interesting to note that the correlation
function of clusters of galaxies is compatible with \eref{power}
once we replace $r_g$ by
$r_c \simeq 15 - 30 \, h^{-1}$ Mpc.  The value of $r_c$ is rather
controversial due to possible selection effects in the compilation of
the catalogues of galaxy clusters \cite{contr}, but however
exceeds $r_g$ by a significant factor.

\subsection{Moments of the cell-counts and scaling}
Let us center a sphere of radius $r$ on a galaxy (labeled by $i$)
and call $n_i(r)$ the number of galaxies in it excluding the
central one. Averaging over the $N$ galaxies of the sample we get the
mean count
\begin{equation}
\langle N \rangle_r \equiv { 1 \over N } \sum_{i=1}^N n_i(r) \, ,
\end{equation}
which provides the correlation integral \eref{coin}.
We shall say that there is {\sl scaling} for the first moment
of the counts of neighbors if
\begin{equation}
\langle N \rangle_r \propto r^{D_2} \, , \label{scald2}
\end{equation}
and the constant exponent $D_2$ is known as {\sl correlation
dimension} \cite{Mar90a}.

In the range of scales where $\xi(r) \gg 1$, equations
\eref{coin}, \eref{power}, \eref{scald2} allow us to derive
a relation between the
exponent of the two-point correlation function $\gamma$ and $D_2$
\begin{equation}
D_2 \simeq 3 - \gamma \, .
\end{equation}
Obviously, in the regime where $\xi(r)$ is of the order of unity, the
previous relation does not hold.

Scaling may be generalized to moments of any order if
\begin{equation}
Z(q,r) = {1 \over N} \sum_{i=1}^N n_i (r)^{q-1} \propto r^{\tau(q)} \,
, \label{scalge}
\end{equation}
with scaling indices $\tau(q)$ independent of $r$ in a suitable
interval. There we define the {\sl generalized dimensions}
$D_q = \tau(q)/(q-1)$. For $q=2$, we recover
the scaling of \eref{scald2}, where $\tau(2)=D_2$.
When the scaling relation \eref{scalge} holds,
we will say that the point distribution has multifractal
character. In a simple fractal $D_q = \mbox{\rm const}$ for all $q$
values, while for a  multifractal set $D_q$ is a decreasing function
of $q$. The meaning of $D_q$ is clear: when $q$ is positive and large,
the denser parts of the point distribution dominate the sums in
\eref{scalge}, while for negative values of $q$ the sums are dominated
by the rarefied regions of the point set.

For $q < 2$, it is usually more convenient to obtain $D_q$ through a
different algorithm. Let us call $r_i(n)$ the radius of the smallest
sphere centered at point $i$ and enclosing $n$ neighbors, in other
words $r_i(n)$ is the distance of point $i$ to its $n$th neighbor.
The exponents $\tau(q)$ are obtained through the relation
\begin{equation}
W(\tau,n) = {1 \over N} \sum_{i=1}^N r_i(n)^{-\tau} \propto n^{1-q} \,
{}.
\end{equation}

For $q=0$, the previous equation provides a way for estimating the
Hausdorff dimension $D_0$. Galaxy samples such as the CfA-I provide
a value of $D_0 \simeq 2.1$, while the correlation dimension of the
same sample is $D_2 \simeq 1.3$ \cite{Mar90b} in the range of scales
$ 1 \le r \le 10 \, h^{-1}$ Mpc. The whole $D_q$ function of a
volume-limited limited subsample of the CfA-I catalogue is illustrated
in Fig. 2 (solid line). Equation (15) has been used for $q \ge
2$, while Equation (16) has been used for $q < 2$.

\subsection{Multisacling}
Beyond $10 \, h^{-1}$ Mpc it is much more difficult to estimate the
galaxy-galaxy two-point correlation function. Nevertheless, integral
quantities such as $C(r)$ can still be estimated with enough
reliability. For $r\le 10 \, h^{-1}$ Mpc, there is also information
about the statistics of the distribution of clusters of galaxies. In
this section, we will provide an explanation work of the clustering of
different objects such as galaxies or clusters within the same
theoretical framework.

If the matter distribution is considered a continuous density field,
we could think of galaxies as being the peaks of the field above some
given threshold. A larger threshold will correspond to clusters of
galaxies. The higher the threshold the richer the galaxy cluster. We
can use the stochastic model shown in Fig. 3$(a)$ to illustrate this
behaviour (for details see \cite{jon92,sil95}). By applying different
density thresholds, which are quite naturally defined in this model,
we obtain the distributions shown in Figs. 3$(b),(c)$. The parameters
of the model were chosen to provide values for $D_0 = 2.0$ and for
$D_2 = 1.3$. The whole $D_q$ function of this model is
illustrated in Fig. 2 (dashed line).

The value of $D_2$ is approximately the slope of
the log-log plot $Z(2,r)$ vs. $r$
shown as a solid line in Fig. 4. After applying the threshold, $Z(2,r)$
stills follows a power law, but with different slope (see dotted and
dashed lines in the figure). In the plot we
can see that the higher the density threshold, the lower is $D_2$.
Multiscaling is a scaling law where the exponent is slowly varying
with the length scale due to the presence of a threshold density
defining the objects \cite{multis}.

We have seen that the observed matter distribution in the Universe
follows some sort of multiscaling behaviour \cite{Mar95}.
If galaxies and clusters of galaxies with increasing richness are
considered as different realizations of the selection of a density
threshold in the mass distribution, the multiscaling argument
implies that the corresponding values of the correlation dimension
$D_2$ must decrease with increasing density.

We shall show the correlation integral for galaxy samples and for
cluster samples in the range [10, 50] $h^{-1}$ Mpc. In this range of
scales $\xi_{gg}(r)$ does not follow a power-law
shape, while $C(r)$ is nicely fitted to a power-law shape.
For galaxies we have analyzed the CfA-I sample, the
Pisces-Perseus sample \cite{PP} and the QDOT-{\it IRAS} redshift survey
\cite{Mar94}.
The cluster samples are the Abell and ACO catalogues \cite{Abell}, the
Edinburgh-Durham redshift survey \cite{ED}, the ROSAT X-ray-selected
cluster sample \cite{Xray} and the APM cluster catalogue \cite{Apm}.

In Fig. 5 we see that three straight lines fit reasonably well the
eight samples analyzed.
All the cluster samples have a correlation integral
well fitted by a power-law with exponent $D_2 \simeq 2.1$.
A value of $D_2 \simeq 2.5$ appears for the optical galaxy catalogues:
the CfA volume-limited sample and the Pisces-Perseus survey.
Finally a value of $D_2 \simeq 2.8$ is obtained for the
QDOT-{\it IRAS} galaxies.
These results probe the multiscaling behaviour of the matter
distribution in the Universe.

The fact that $D_2$ for {\it IRAS} galaxies is larger than for
optical samples indicates that {\it IRAS} galaxies are less
correlated than optical galaxies; this is nicely interpreted if
optical galaxies correspond to higher peaks of the density field.
Clusters of galaxies have stronger correlations than galaxies,
corresponding to the highest peaks of the background matter
density.

\section{Conclusions}
We have given a short introduction to different aspects of the
characterization of the observed surveys of galaxy clustering by means
of different statistical techniques. Redshift surveys, when considered
as point processes, have peculiar features which can be expressed
through statistical tools. Obscuration by
dust in our own Galaxy, truncation in luminosity and the
use of selection functions for flux-limited samples have been
discussed in some detail. The analysis of clumpiness
is often done in redshift space, which has important distortions
when compared to real space.
Morphological and luminosity segregation is an important
clue for testing galaxy formation theories. It is interesting
to consider the galaxy distribution as a marked point process.
We have illustrated the relationship of the two-point correlation
function to other statistical quantities such as the intensity
functions or cumulant quantities such as the correlation integral.
The multifractal nature of the matter distribution comes from the
scaling of the moments of the cell-counts.
We have introduced the concept of multiscaling to provide a neat
scheme for the explanation of the clustering of galaxies of different
kinds and clusters with different richness. In this context, we have
shown how the correlation dimension $D_2$ attains specific values
for each kind of cosmic object, being a clear measure of their
clustering.

\ack

The author wishes to thank his collaborators S. Borgani, P. Coles,
S. Paredes and M.J. Pons for countless discussions on all
aspects of the clustering phenomenon. This work is partially
supported by the EC Human Capital and Mobility Programme network
(Contract ERB CHRX-CT93-0129), by the project number GV-2207/94
of the Generalitat Valenciana and by the Instituci\'o Valenciana
d'Estudis i Investigaci\'o.

\smallskip

\newpage
\section*{Figure captions}

{\bf Figure 1.} The selection function of the QDOT-{\it IRAS} redshift
survey.

{\bf Figure 2.} The generalized dimensions $D_q$ for a volume-limited
subsample of the CfA-I catalog (solid line) and for the multifractal
model shown in Fig. 3 (dashed line).

{\bf Figure 3.} A multifractal stochastic model for the galaxy
distribution $(a)$. In $(b)$ and $(c)$ we see the same model after
applying increasing density thresholds.

{\bf Figure 4.} The function $Z(2,r)=C(r)/n$ for the model shown
in Fig. 3. The slope $D_2$ is lower for samples with higher density
threshold.

{\bf Figure 5.} The correlation integral for different galaxy and
cluster samples (reproduced with permission from Martinez et
al. {\it Science} {\bf 269}, 1245. Copyright 1995 American
Association for the Advancement of Science).

\end{document}